# Boosting line intensity map signal-to-noise ratio with the Ly-$\alpha$ forest cross-correlation

Mahdi Qezlou[⋆](ORCID),[1,2] Simeon Bird[ORCID],[1] Adam Lidz,[3] Guochao Sun,[4] Andrew B. Newman,[2] Gwen C. Rudie,[2] Yueying Ni,[5,6] Rupert Croft[5] and Tiziana Di Matteo[ORCID][5,7]

[1]*Department of Physics and Astronomy, University of California Riverside, 900 University Ave, Riverside, CA 92521, USA*
[2]*The Observatories of the Carnegie Institution for Science, 813 Santa Barbara Street, Pasadena, CA 91101, USA*
[3]*Center for Particle Cosmology, Department of Physics and Astronomy, University of Pennsylvania, Philadelphia, PA 19104, USA*
[4]*California Institute of Technology, 1200 E. California Blvd., Pasadena, CA 91125, USA*
[5]*McWilliams Center for Cosmology, Department of Physics, Carnegie Mellon University, Pittsburgh, PA 15213, USA*
[6]*Harvard–Smithsonian Center for Astrophysics, 60 Garden Street, Cambridge, MA 02138, USA*
[7]*NSF AI Planning Institute for Physics of the Future, Carnegie Mellon University, Pittsburgh, PA 15213, USA*

**ABSTRACT**
We forecast the prospects for cross-correlating future line intensity mapping (LIM) surveys with the current and future Ly-$\alpha$ forest measurements. Using large cosmological hydrodynamic simulations, we model the emission from the CO rotational transition in the CO Mapping Array Project LIM experiment at the 5-yr benchmark and the Ly-$\alpha$ forest absorption signal for extended Baryon Acoustic Oscillations (BOSS), Dark energy survey instrument (DESI), and Prime Focus multiplex Spectroscopy survey (PFS). We show that CO × Ly-$\alpha$ forest significantly enhances the detection signal-to-noise ratio (S/N) of CO, with up to 300 per cent improvement when correlated with the PFS Ly-$\alpha$ forest survey and a 50–75 per cent enhancement with the available eBOSS or the upcoming DESI observations. This is competitive with even CO × spectroscopic galaxy surveys. Furthermore, our study suggests that the clustering of CO emission is tightly constrained by CO × Ly-$\alpha$ forest due to the increased sensitivity and the simplicity of Ly-$\alpha$ absorption modelling. Foreground contamination or systematics are expected not to be shared between LIM and Ly-$\alpha$ forest observations, providing an unbiased inference. Ly-$\alpha$ forest will aid in detecting the first LIM signals. We also estimate that [C II] × Ly-$\alpha$ forest measurements from Experiment for Cryogenic Large-Aperture Intensity Mapping and DESI/eBOSS should have a larger S/N than planned [C II] × quasar observations by about an order of magnitude.

## 1 INTRODUCTION

Line intensity mapping (LIM) experiments have emerged as a powerful technique to study the interstellar medium (ISM) and the diffuse gas within the circumgalactic medium or intergalactic medium (IGM) by observing the aggregate atomic/molecular line emission (Visbal & Loeb 2010; Kovetz et al. 2017; Bernal & Kovetz 2022). This observing strategy complements resolved observations with current or future flagship observatories such as *JWST* (Gardner et al. 2006), Roman (Spergel et al. 2013), HabEx, LUVOIR (The LUVOIR Team 2019), and Origins telescopes (Meixner et al. 2019). LIM data are sensitive to the total line emission, including faint galaxies below the magnitude limits of even flagship observatories (Kovetz et al. 2019). LIM experiments can constrain the distribution of cold gas across cosmic time, which serves as the fuel for star formation in galaxies (Keating et al. , 2020; Sun et al. 2021; Chung et al. 2022; Sun 2022). LIM observations may also shed light on the role of early galaxy formation in reionizing the neutral gas (Lidz et al. 2009, 2011; Gong et al. 2012; Kannan et al. 2022; Sun et al. 2022). Additionally, LIM experiments measure the total integrated emission even from the faintest sources, making it easier to probe the large volumes accessible at higher redshifts compared to spectroscopic galaxy surveys at lower redshifts. This feature positions LIM to address questions in cosmology, such as constraining dark matter models (Creque-Sarbinowski & Kamionkowski 2018) and inflationary models of the early Universe (Moradinezhad Dizgah, Keating & Fialkov 2019).

LIM has recently expanded from a focus on the 21-cm emission of atomic hydrogen to measurements of other atomic or molecular line emissions, particularly at cosmic noon $z \sim 2$–3. For example, the CO Mapping Array Project (COMAP; Cleary et al. 2022) started a 5-yr Pathfinder programme to observe the CO(1–0) rotational transition at $z = 2.4$–3.4 and CO(2–1) at $z = 6$–8 over 12.3 deg$^2$. In the next phase, COMAP End of Reionization programme (EOR) will obtain a higher sensitivity to the same emission lines while expanding the detection to CO(1–0) emission at $z = 4.8$–8.6 (Breysse et al. 2022). The Experiment for Cryogenic Large-Aperture Intensity Mapping (EXCLAIM), on the other hand, will observe the [C II] line emission at $z = 2.5$–3.5 and CO rotational lines at $z < 1$ over 305 deg$^2$ of the sky, rich with large extragalactic data archives such as Sloan Digital Sky Survey (SDSS), Baryon Oscillation Spectroscopic

⋆ E-mail: mahdi.qezlou@email.ucr.edu

Survey (BOSS; Ahumada et al. 2020), and Hyper Supreme Cam (HSC) photometric galaxy survey (Aihara et al. 2022).

However, it is challenging to detect the LIM signal owing to detector noise, and foreground contamination from our own galaxy, as well as extragalactic sources, including interloper emission lines in some cases. Interloper lines are emissions at different rest-frame frequencies but redshifted to the observed channel, which can lead to misinterpretations of the signal. One potential solution to overcome this issue is to cross-correlate the LIM signal with other known tracers of large-scale structure, which can improve the sensitivity of the measurement. For example, Furlanetto & Lidz (2007) show that the sensitivity of a 21-cm probe of the neutral hydrogen at the epoch of reionization improves by a factor of several when cross-correlated with a galaxy redshift survey. Similar cross-correlations have also been used to constrain the 21 cm × galaxy (Wolz et al. 2022) and Ly-$\alpha$ emission × absorption (Renard et al. 2021) at lower redshifts. Pullen et al. (2022) predict that the signal in the EXCLAIM experiment can be detected through the cross-correlation between [C II] emission, at $z = 2.5$–3.5, and the quasars in Stripe 82 of eBOSS. The first constraints on the small-scale power of emission from the CO molecule's rotational transitions at $z \sim 3$ have been obtained by cross-correlating with galaxies (Keating et al. 2015, 2016b; Keenan, Keating & Marrone 2022). Nevertheless, Chung et al. (2019) demonstrate that the constraining power of the CO × galaxy is limited by the mass completeness and redshift uncertainty of the galaxy sample.

The frequent resonant scattering of the light by the neutral hydrogen within the IGM leads to distinctive absorption features in the spectra of background galaxies. This large absorption field, commonly called the Ly-$\alpha$ forest, traces the large-scale distribution of the underlying matter with $\sim$Mpc resolution. A dense collection of these spectra can make a tomographic map of the large-scale structure (Lee et al. 2014; Horowitz et al. 2022), which provides a unique opportunity to study the relationship between the galaxy properties and their environment (Momose et al. 2022; Newman et al. 2022; Qezlou et al. 2022; Dong et al. 2023). The largest such tomographic survey with a high density of sightlines covers around $\sim$1.7 deg$^2$ of the sky at $z = [2.2–2.8]$ through spectroscopy of the Lyman-break galaxies (LBGs) and QSOs (Newman et al. 2020). However, the upcoming Prime Focus Spectrograph (PFS) will map $\sim$10 deg$^2$ within the same cosmic time window, similar to the observed volume of the planned LIM pathfinders like COMAP (Chung et al. 2022). Quasar-based Ly-$\alpha$ forest surveys such as eBOSS (Ravoux et al. 2020) and DESI (Chaussidon et al. 2023) map larger volumes ($\sim$220 and 14 000 deg$^2$, respectively), which facilitate overlap with upcoming LIM experiments; however, the sparsity of the sources results in lower transverse resolution.

In this study, we investigate the prospects of detecting molecular line emission through cross-correlation with the Ly-$\alpha$ forest, particularly the full data release of COMAP experiment (Chung et al. 2022). In Section 2, we introduce the hydrodynamic simulation we use. Section 3 describes how we generate synthetic observations from the simulated data. Section 4 details the summary statistics we use and how we model the noise. Our measurement forecasts for these statistics are provided in Section 5. We discuss the implications of our findings and the robustness of the results in Section 6. Finally, the main findings are summarized in Section 7.

## 2 SIMULATION

We create mock observations using the cosmological simulation ASTRID (Bird et al. 2022; Ni et al. 2022). ASTRID is run using MP-GADGET, a smoothed particle hydrodynamics (SPH) code and a modified version of GADGET-3 used for the BlueTides simulation. The simulation initially contains $2 \times 5500^3$ dark matter and gas particles within a periodic box of 250 $h^{-1}$ cMpc. The gravity solver is based on a Tree-PM algorithm, which divides the force computation between a long-range particle mesh and a short-range hierarchical tree. The multiphase prescription of star formation from Springel & Hernquist (2003) is adopted along with radiative and metal cooling of the gas particles (Katz, Weinberg & Hernquist 1996; Vogelsberger et al. 2014). A small correction to the star formation is applied to account for molecular hydrogen formation (Krumholz & Gnedin 2011). Self-shielding of the dense gas is modelled using the fitting function of Rahmati et al. (2013). Newly formed star particles source hydrodynamically decoupled galactic winds, which recouple using a density change or time threshold. The simulation places the supermassive black hole (SMBH) seeds in haloes with a friend-of-friend halo mass larger than $5 \times 10^9 \, h^{-1} \, M_\odot$ and a stellar mass larger than $2 \times 10^6$ by converting the densest gas particle in the halo. The SMBH seed mass is drawn from a power-law distribution between $3 \times 10^4$ and $3 \times 10^5 \, h^{-1} \, M_\odot$. SMBHs are kept near their host galaxy centre with an effective dynamical friction force following Chen et al. (2022). ASTRID implements black hole accretion following a Bondi–Hoyle–Lyttleton-like prescription (Di Matteo, Springel & Hernquist 2005) and thermal black hole feedback (active galactic nucleus) with an efficiency of 5 per cent (Chen et al. 2022). We refer to Ni et al. (2022) for more details of the SMBH model used in ASTRID.

In ASTRID, the patchy reionizations of hydrogen and helium occurring at $z > 6$ and $z > 2.8$, respectively, have significant implications for the IGM. Hydrogen reionization is modelled using a pre-computed reionization redshift map based on the large-scale smoothed overdensity, as described in Battaglia et al. (2013). On the other hand, helium reionization is implemented by randomly placing large bubbles at the probable location of quasars in massive haloes, as outlined in Upton Sanderbeck & Bird (2020). The subhaloes of dark matter used in this study are obtained using the SUBFIND algorithm (Springel et al. 2001) in post-processing, and their corresponding subhalo masses are denoted as $M_h$.

## 3 MOCK OBSERVATIONS

This section provides a brief overview of the methods used to generate mock observations and how these simulations can represent both current and upcoming observational surveys. Refer to Table 1 for the details of the surveys considered in this work.

### 3.1 Mock Ly-$\alpha$ tomography

We produce artificial Ly-$\alpha$ absorption sightlines using the fast FAKE_SPECTRA[1] PYTHON package (Bird et al. 2015; Bird 2017; Qezlou et al. 2022). Each gas particle is treated as a separate absorber, and the absorption from all gas particles along the line of sight is summed to calculate the absorption spectra. Absorption from a single gas particle is computed by convolving a Voigt profile with the particle's density kernel. The internal physical quantities in each SPH particle are smoothed using a quintic spline kernel.[2] At $z \sim 2.5$, the neutral hydrogen fraction is computed by solving a rate network assuming ionization equilibrium between the uniform ionizing ultraviolet (UV) background and the recombination rates from Katz et al. (1996). The observed mean flux imposed by the uniform UV background (Faucher-Giguère et al. 2008) is enforced in the simulated spectra by scaling the overall optical depth, similar to Rauch et al. (1997), Croft et al. (1998), and Qezlou et al. (2022). The simulated forest recovers the 1D flux power spectrum measured by the SDSS Data Release 14 (DR14; Chabanier et al. 2019) at the 10 per cent level. The final Ly-α absorption map generated for the power spectrum calculations is on a uniform grid with side length of $250\,h^{-1}$ ckpc.

---

[1] https://github.com/sbird/fake_spectra
[2] A top-hat for TNG300 simulations, see Appendix A.

**Table 1.** Summary of the key parameters in COMAP-Y5, Ly-α forest and galaxy surveys used in this work.

| LIM Survey | area [deg$^2$] | $z_{range}$ | timeline | Ref | |
|---|---|---|---|---|---|
| COMAP-Y5 | 12 | 2.4–3.4 | From 2022–2027 | Cleary et al. (2022) | |
| Ly-α Forest Survey | $d_\perp$ | area [deg$^2$] | $z_{range}$ | timeline | Ref |
| Ly-α, eBOSS | 13 | 220 | 2.2–3.0 | completed | Ravoux et al. (2020) |
| Ly-α, DESI | 10 | 14 000 | 2.2–3.0 | started | Chaussidon et al. (2022) |
| Ly-α, PFS-bright | 3.7 | 12.3 | 2.2–2.6 | From 2024 to 2029 | Greene et al. (2022) |
| Ly-α, PFS-Faint | 2.5 | 12.3 | 2.2–2.6 | From 2024 to 2029 | Greene et al. (2022) |
| Spectroscopic Galaxy Survey | Targets (sampling rate) | $\frac{\sigma_z}{(1+z)}$ | area [deg$^2$] | $z_{range}$ | timeline | Ref |
| PFS | 10 870 (34 %) | $7 \times 10^{-4}$ | 12.3 | 2.2–3.5 | From 2024–2029 | Takada et al. (2014); Greene et al. (2022) |
| Photometric Galaxy Survey | Selection | $\frac{\sigma_z}{(1+z)}$ | area [deg$^2$] | $z_{range}$ (for this work) | timeline | Ref |
| HSC- U+grizy+ YJHK (10-bands) | i < 25 or $M_h > 10^{11.51}\,h^{-1}\,M_\odot$ | $2 \times 10^{-2}$ | 5.5 | 2.0–3.0 | completed | Desprez et al. (2023) |
| HSC- U+grizy+ YJHK (10-bands) | i < 26 or $M_h > 10^{11.21}\,h^{-1}\,M_\odot$ | $3 \times 10^{-2}$ | 5.5 | 2.0–3.0 | completed | Desprez et al. (2023) |
| HSC- U+grizy (6-bands) | i < 25 or $M_h > 10^{11.31}\,h^{-1}\,M_\odot$ | $4 \times 10^{-2}$ | 18.6 | 2.0–3.0 | completed | Desprez et al. (2023) |
| HSC- U+grizy (6-bands) | i < 26 or $M_h > 10^{11.01}\,h^{-1}\,M_\odot$ | $6 \times 10^{-2}$ | 18.6 | 2.0–3.0 | completed | Desprez et al. (2023) |

In recent years, significant progress has been made in mapping the IGM using Ly-α absorption tomography. We consider two classes of Ly-α survey with mean background source separations of $d_\perp = 10$–13 or 2.5–3.7 $h^{-1}$ cMpc. The eBOSS survey, covering 220 deg$^2$ of the sky at $z = 2.2$–3.0, achieved a mean sightline separation of $d_\perp \sim 13\,h^{-1}$ cMpc using QSO spectra from the SDSS DR16 Stripe 82 region (Ravoux et al. 2020). Additionally, the DESI survey has started a 5-yr programme to observe QSOs at $z > 2.1$ over 14 000 deg$^2$ of the sky, achieving a mean separation of $d_\perp \sim 10\,h^{-1}$ cMpc (Chaussidon (2023) ). For higher spatial resolution, spectroscopy of fainter sources like LBGs over a smaller footprint has been pursued by COSMOS Lyman-Alpha Mapping And Tomography Observations (CLAMATO), which maps 0.2 deg$^2$ of sky within $z = 2.05$–2.55, achieving a mean transverse resolution of $d_\perp \sim 2.5\,h^{-1}$ cMpc (Lee et al. 2018; Horowitz et al. 2022). The largest high-resolution tomography is performed with Lyman-Alpha Tomography IMACS Survey (LATIS), which maintains similar mean transverse resolution across 13 times larger volume at $z = 2.2$–2.8 (Newman et al. 2020, 2022; Qezlou et al. 2022). A tomography survey with the multiplex spectroscopy on the Prime Focus multiplex Spectroscopy (PFS) instrument is planned to obtain similar mapping quality, i.e. $d_\perp = 2.5$–3.7 $h^{-1}$ cMpc, over ∼12.3 deg$^2$ (Greene et al. 2022). This survey is large enough to potentially fully overlap with a Pathfinder line intensity survey and is set to begin in 2024. However, the spatial resolution of the cross-correlated signal with CO emission in such dense tomographies will also be limited by the beam size and channel width of the COMAP-Y5 Pathfinder, i.e. $(d_\parallel, d_\perp) \sim (2.64, 4.66)\,h^{-1}$ cMpc (Chung et al. 2022). For our mock observations, we adopt a realistic spectral signal-to-noise ratio (S/N) of 2 per Å for all sources, consistent with typical Ly-α forest data from surveys such as eBOSS (Lee et al. 2013), CLAMATO (Lee et al. 2018), and LATIS (Newman et al. 2020). As noted by McQuinn & White (2011), longer exposure for individual spectra does not significantly improve the survey's sensitivity to the autocorrelation of the forest data. In Table 1, we summarize the details of the surveys considered in this work.

### 3.2 Mock galaxy surveys

To generate a mock galaxy survey for our analysis, we adopt a method similar to previous studies on CO × galaxy surveys (Li et al. 2016; Chung et al. 2019) and 21-cm signal × galaxies (Furlanetto & Lidz 2007). We construct a galaxy density map at $z \sim 2.5$ by displacing subhaloes based on their peculiar velocities along the line of sight and then mapping them on to a uniform fine grid with a side length of $250\,h^{-1}$ ckpc using the cloud-in-cell kernel. The power spectrum of this density map in redshift space is the signal in our analysis.

In Section 4, we incorporate the effects of observed galaxy redshift uncertainties by accounting for them in the noise power spectrum, following the methodology developed in Furlanetto & Lidz (2007) and Chung et al. (2019). We consider two classes of galaxy redshift surveys: *photometric* and *spectroscopic*. An example of a wide and deep *photometry* survey at $z = 2$–3 is the deep layer of Hyper Suprime-Cam Subaru Strategic Program (Aihara et al. 2022), which imaged 36 deg$^2$ in five broad-band filters of *grizy*. The typical redshift uncertainties at the relevant redshifts, ignoring the outliers, are estimated to be $\sigma_z/(1+z) = 0.09$ by Nishizawa et al. (2020). Further imaging of 18.6 deg$^2$ of these galaxies with additional filters in the *u* band by the CLAUDES project (Sawicki et al. 2019) reduces the redshift uncertainties to $\sigma_z/(1+z) \simeq 0.05$. Moreover, observations in auxiliary near-infrared (IR) bands (*YJHK*) can reduce the uncertainties to $\sigma_z/(1+z) = 0.03$ for a smaller subset of 5.5 deg$^2$ of these galaxies (Desprez et al. 2023). The *photoz* quality of the `SourceExtractor`[3] catalogue provided in Desprez et al. (2023) degrades for fainter galaxies; therefore, we consider two subsets with magnitude cuts of $i < 25$ or $i < 26$ (similar to those in Nishizawa et al. 2020) for each of the 6-band (*ugrizy*) or 12-band (*ugrizy + YJHK*) observed catalogues. We match the source densities of these catalogue subsets with the abundance of the subhaloes in ASTRID more massive than a halo mass ($M_h$) threshold, resulting in slightly different mass completeness for each subset, as summarized in Table 1. An accurate abundance-matching technique, however, requires a better understanding of the selection function of the observed galaxy samples. We also consider an upcoming medium-resolution *spectroscopic* galaxy survey over 12.3 deg² at $z = 2.2-3.5$ within the PFS galaxy evolution project (Greene et al. 2022). The redshift uncertainties of such a galaxy survey at $z \sim 2.5$ are expected to be about $\sigma_z/(1 + z) = 7 \times 10^{-4}$, which is required for cosmological studies (Takada et al. 2014). Greene et al. (2022) estimates $\sim 10\,870$ available targets in the 1.3-deg² field of view when observing in the faint mode, of which $\sim 34$ per cent will be targeted. We build mock spectroscopic galaxy surveys by random sampling from haloes with $M_h > 10^{11.9}\,h^{-1}\,\mathrm{M}_\odot$, following the same subhalo abundance-matching technique.

### 3.3 Mock CO LIM

We model the CO emission from galaxies using a power-law scaling relation between line luminosity and average star formation in haloes. To this end, we adopt the prior provided by the COMAP Early Science project (Chung et al. 2022), which links the scaling relation between CO(1–0) luminosity ($L_\mathrm{CO}$) and star formation rate (SFR) to $M_h$ through a double power-law parametrization modified from Padmanabhan (2018). This uses an analytical fit between the average SFR and halo mass proposed by the UniverseMachine framework (Behroozi et al. 2019):

$$\frac{L'_\mathrm{CO}}{K\,\mathrm{km\,s^{-1}\,pc^2}} = \frac{C}{(M_h/M)^A + (M_h/M)^B}$$
$$\frac{L_\mathrm{CO}}{L_\odot} \sim \mathrm{LogNorm}(\mu = 4.9 \times 10^{-5} L'_\mathrm{CO}, \sigma = \sigma)\,, \quad (1)$$

where LogNorm indicates a lognormal distribution and $A, B, C, M$, and $\sigma$ are the model parameters, which are broadly constrained by the observations from COLDz (Riechers et al. 2019) and COPSS (Keating et al. 2016b). We adopt the fiducial model parameters from table 5 of Chung et al. (2022) and discuss the sensitivity of our results to variations in these parameters in Section 6. Although the ASTRID simulation accurately reproduces the observed SFR (Bird et al. 2022), we do not attempt to build a new CO emission model based on these predictions, aiming to be consistent with the COMAP analysis. We postpone this exploration to future work.

We construct the CO temperature map by integrating the emission from all subhaloes ($M_\mathrm{SUBFIND} > 10^7\,h^{-1}\,\mathrm{M}_\odot$) within voxels of side length $250\,h^{-1}\,\mathrm{ckpc}$ using a cloud-in-cell interpolation. The CO luminosity to temperature conversion ($L_\mathrm{CO}$–$T_\mathrm{CO}$) is done using the standard conversions in units of $\mu$K, as described in appendix B.1 in Chung et al. (2019):

$$T_\mathrm{CO} = 3.1 \times 10^4\,\mu\mathrm{K}\,(1+z)^2 \left(\frac{\nu_\mathrm{rest}}{\mathrm{GHz}}\right)^{-3},$$
$$\times \left(\frac{H(z)}{\mathrm{km\,s^{-1}\,Mpc^{-1}}}\right) \left(\frac{L_\mathrm{CO,vox}}{L_\odot}\right) \left(\frac{V_\mathrm{vox}}{\mathrm{Mpc}^3}\right)^{-1}, \quad (2)$$

---
[3] https://www.clauds.net/available-data

where $L_\mathrm{CO,vox}$ is the total CO luminosity in each voxel with a volume of $V_\mathrm{vox}$. COMAP Pathfinder experiment (Cleary et al. 2022) observes the CO(1–0) rotational transition (rest-frame frequency $\nu_\mathrm{rest} = 115.27$ GHz) at redshifts between 2.4 and 3.4 in three fields of 4 deg² each. At the 5-yr mark of the Pathfinder experiment, the system temperature fluctuation amplitude is expected to be $\sim 17.8\,\mu$K per map voxel, with a voxel size of $(d_\parallel, d_\perp) = (2.64, 4.66)\,h^{-1}\,\mathrm{cMpc}$ at redshift of $z \simeq 2.5$ (Chung et al. 2022). We account for the instrumental noise and finite angular/spectral resolution in our model of the CO noise power spectrum (Section 4). In our analysis, we assume that COMAP-Y5-like volume fully overlaps with the other auxiliary surveys, either galaxy or Ly-$\alpha$ forest data.

## 4 STATISTICS

The primary summary statistic considered in this work is the spherically averaged power spectrum. To estimate the signal power spectrum, we perform a fast Fourier transform of the noiseless signal on a uniform fine grid with a side length of $250\,h^{-1}$ ckpc. However, due to asymmetric uncertainties along and transverse to the sightline, we first calculate the power spectrum in $k$–$\mu$ space using the following equation:

$$\sigma_{P_\mathrm{A}}(\mathbf{k}, \mu) = \frac{P_\mathrm{A,s}(\mathbf{k}, \mu) + P_\mathrm{A,n}(\mathbf{k}, \mu)/W_\mathrm{A}^2(\mathbf{k}, \mu)}{\sqrt{N_\mathrm{m}(\mathbf{k}, \mu)}}\,, \quad (3)$$

where $\mu$ is the cosine of the angle between the line of sight and $k$ the wavevector $\mathbf{k}$. Here, $P_\mathrm{A,s}$ and $P_\mathrm{A,n}$ are the signal and noise power spectra for any mock observation $A$, respectively. $N_\mathrm{m}(k, \mu)$ is the mode count in each bin, and $W_\mathrm{A}^2(k, \mu)$ is the signal attenuation term due to the finite angular and spectral resolution of the survey. The first and second terms in equation (3) account for sample variance in the limited observed volume and the noise contribution in each survey, respectively.

For CO LIM, we adopt the noise model and survey parameters from the COMAP Pathfinder survey (Chung et al. 2019, 2022). Specifically, the noise power spectrum is given by

$$P_\mathrm{CO,noise}(\mathbf{k}, \mu) = \sigma_n^2 V_\mathrm{vox,COMAP},$$
$$W_\mathrm{CO}^2(\mathbf{k}, \mu) = e^{-k^2 \sigma_\perp^2} e^{-\mu^2 k^2 (\sigma_\parallel^2 - \sigma_\perp^2)}\,, \quad (4)$$

where $\sigma_n$ is the noise temperature in each voxel of volume $V_\mathrm{vox,COMAP}$ and $(\sigma_\parallel, \sigma_\perp)$ are the voxel sizes parallel and transverse to the sightline (Section 3.3).

For Ly-$\alpha$ tomography, we estimate the uncertainties in the 3D power spectrum following McQuinn & White (2011). They optimize the S/N in the power spectrum for a set of weights that quantify the contribution of each spectrum to the total signal. The noise power spectrum is then given by

$$P_{\mathrm{Ly}\alpha,n}(k, \mu) = P_\mathrm{los}(k_\parallel)/\bar{n}_\mathrm{2D,eff}, \quad (5)$$

where $\bar{n}_\mathrm{2D,eff}$ is a noise-weighted projected density of the background source galaxies, given by

$$\bar{n}_\mathrm{2D,eff} = \frac{1}{A} \sum_{i=1}^{N} \nu_i\,, \quad (6)$$

$$\nu_i = \frac{P_\mathrm{los}(k_\parallel)}{P_\mathrm{los}(k_\parallel) + P_{\mathrm{N},i}(k_\parallel)}\,. \quad (7)$$

$A$ is the survey area and $P_{\mathrm{N},i}$ is the 1D noise power spectrum of the $i$th source. In this work, we assume $P_{\mathrm{N},i}(k_\parallel)$ to be white noise with an S/N per angstrom $= 2$, typical of these surveys. Specifically, we have

$$P_{\mathrm{N},i}(k_\parallel) = (<F>/(\mathrm{S/N}))^2 \Delta X\,. \quad (8)$$

$<F>$ is the mean absorption flux in the forest and $\Delta X = \lambda_{Ly\alpha} H(z) h/c$ is the conversion factor from Å to $h^{-1}$ cMpc. Since this estimate is not convolved with any instrumental beam, we do not apply deconvolution to the noise power in equation (3), i.e. $W_{Lya}(k, \mu) = 1$. We note that the proposed noise power spectrum in equation (5) accounts only for the Poisson noise of the background sources. The clustering of the background sources becomes important only for tomography surveys with a higher sightline density than we consider here, $d_\perp \ll 2.5 \, h^{-1}$ cMpc (McQuinn & White 2011). This is further discussed in Section 6.

In galaxy surveys, the noise power spectrum depends on the number density of the galaxies, $n_{3D}$, amplified by the redshift uncertainty:

$$P_{\text{Gal,noise}} = 1/n_{3D},$$
$$W_{\text{gal}}^2(k, \mu) = e^{-\mu^2 k^2 \sigma_\parallel^2}, \quad (9)$$

where $\sigma_\parallel$ is the spatial resolution that is related to the redshift uncertainty, $\Delta_z = \sigma_z/(1+z)$, through

$$\sigma_\parallel = \frac{c \, \sigma_z}{H(z)} = \frac{c(1+z)}{H(z)} \Delta_z. \quad (10)$$

The uncertainties in the cross-power spectra between the CO signal and other tracers, such as Ly-$\alpha$ or galaxies, are estimated as

$$\sigma_{P_{CO \times A}}^2(\mathbf{k}, \mu) = \frac{\sigma_{P_{CO}}(\mathbf{k}, \mu) \times \sigma_{P_A}(\mathbf{k}, \mu)}{2} + \frac{P_{CO \times A}^2(\mathbf{k}, \mu)}{2 N_m(\mathbf{k}, \mu)}. \quad (11)$$

$A$ represents either an Ly-$\alpha$ or a galaxy survey. The uncertainties in the $k$-bins of the spherically averaged power spectra are then computed by summing the noise power over $\mu$-bins in inverse quadrature as shown in Furlanetto & Lidz (2007):

$$\sigma_{P_A}^{-2}(k) = \sum_\mu \sigma_{P_A}^{-2}(\mathbf{k}, \mu). \quad (12)$$

The S/N at each $k$-bin and the total S/N are defined as

$$\frac{S}{N} = \left[ \sum_k \left( \frac{S}{N}(k) \right)^2 \right]^{1/2} = \left[ \sum_k \left( \frac{P_{\text{noiseless}}(k)}{\sigma_P(k)} \right)^2 \right]^{1/2}. \quad (13)$$

The intrinsic correlation of the CO signal with other tracers, i.e. Ly-$\alpha$ forest or galaxies, is summarized by the cross-correlation coefficient between the noiseless fine-gridded signals:

$$r(k) = \frac{P_{CO \times A}(k)}{\sqrt{P_{CO}(k) P_A(k)}}. \quad (14)$$

## 5 RESULTS

This section presents our findings on the cross-correlation signal between various current or planned Ly-$\alpha$ forest surveys and the COMAP-Y5 intensity map. We adopt the mean of the Gaussianized prior provided in table 5 of Chung et al. (2022) as our fiducial set of CO emission model parameters. Based on the presented forecasts in this section, the cross-correlation of Ly-$\alpha$ tomography with LIM is expected to provide a competitive alternative to the cross-correlation with future galaxy redshift surveys for increasing the sensitivity to the CO signal. We assume that all auxiliary surveys will fully overlap with the COMAP-Y5 volume. For a more detailed discussion and interpretation of the results, please refer to Section 6.

### 5.1 Signal to noise

Fig. 1 illustrates the simulated signals of COMAP-Y5 cross-correlated against Ly-$\alpha$ forest or conventional photometric and spectroscopic galaxy surveys. In the left-hand panel, two such noiseless signals are compared, that is, an Ly-$\alpha$ tomography map with a spatial resolution of $(250 \, h^{-1} \, \text{ckpc})^3$ and a few galaxy surveys with perfect redshift estimation and different mass completeness. Galaxies are better indicators of CO emission on smaller scales compared to Ly-$\alpha$ absorption. This could be because the CO lines originate from the ISM within galaxies, whereas Ly-$\alpha$ absorption traces larger scales in the IGM. On large scales with $k < 0.1 \, h \, \text{Mpc}^{-1}$, the forest becomes less correlated with the underlying matter density due to He II reionization modelled in ASTRID (Bird et al. 2022). He II reionization causes extra absorption on scales larger than $30 \, h^{-1}$ cMpc around the densest regions (McQuinn et al. 2009, 2011; Gontcho A Gontcho, Miralda-Escudé & Busca 2014; Pontzen 2014; Pontzen et al. 2014). For a detailed discussion on the He II reionization effect, refer to Appendix A.

Once realistic completeness and noise models are accounted for, Ly-$\alpha$ tomography surveys become competitive. The middle panel in Fig. 1 compares the forecast S/N for various surveys with an observed volume similar to COMAP-Y5 Pathfinder. The simulated volume is scaled to match the COMAP-Y5 Pathfinder by modifying the term $\sqrt{N_{\text{modes}}} \propto \sqrt{V}$ in equation (3). This scaling is justified since ASTRID's volume has sufficient modes for power spectrum calculations on the scales to which COMAP-Y5 Pathfinder is most sensitive, i.e. $k = [0.05 – 0.6] \, h \, \text{Mpc}^{-1}$ (Ihle et al. 2022). The observed signal beyond this range is significantly contaminated and down-weighted during the pre-processing of the COMAP observations (Foss et al. 2022).

In Fig. 1, the total forecasted S/N is shown in the right-hand panel, estimated using equation (13). The sensitivity to the cross-correlated CO emission with a dense Ly-$\alpha$ tomography, where $d_\perp = 2.5–3.7 \, h^{-1}$ cMpc, is comparable to the cross-correlated field with a medium-resolution spectroscopic galaxy survey. Furthermore, coarse tomography surveys with $d_\perp = 10–13 \, h^{-1}$ cMpc show better sensitivity enhancement than most typical photometric galaxy surveys. The next section discusses the implications of a higher S/N for characterizing the clustering of cosmological line emission.

### 5.2 Forecast parameter inference

To forecast the inference on the clustering power of the CO emission, we model the simulated signal as biased tracers of the linear matter fluctuations in physical space:

$$\hat{P}_{CO} = (\langle T_{CO} \rangle b_{CO})^2 \, P_m(k) + P_{\text{shot,CO}}, \quad (15)$$

$$\hat{P}_{\text{Gal}} = b_{\text{Gal}}^2 \, P_m(k) + P_{\text{shot,gal}}, \quad (16)$$

$$\hat{P}_{\text{Lya}} = b_{\text{Lya}}^2 \, P_m(k), \quad (17)$$

$$\hat{P}_{CO \times \text{Gal}} = b_{\text{Gal}} \langle T_{CO} \rangle b_{CO} \, P_m(k) + P_{\text{shot,CO} \times \text{Gal}}, \quad (18)$$

$$\hat{P}_{CO \times \text{Lya}} = b_{\text{Lya}} \langle T_{CO} \rangle b_{CO} \, P_m(k). \quad (19)$$

These equations do not account for redshift-space distortions, even though it is present in the simulated signal. The CO × Galaxy survey model requires five parameters, namely $\langle T_{CO} \rangle b_{CO}$, $P_{\text{shot,CO}}$, $b_{\text{gal}}$, $P_{\text{shot,Gal}}$, and $P_{\text{shot,CO} \times \text{Gal}}$, while the CO × Ly-$\alpha$ tomography model requires three parameters, namely $\langle T_{CO} \rangle b_{CO}$, $P_{\text{shot,CO}}$, and $b_{\text{Lya}}$. Shot-noise terms are necessary for modelling the auto galaxy, CO, or their cross-power spectra due to the discrete nature of sources. The $P_{\text{shot,CO} \times \text{Gal}}$ term contains exclusive information on the CO emission strength of the sample galaxies in the cross-correlated survey compared to other shot-noise terms (Bernal & Kovetz 2022). In contrast, due to the continuum nature of the H I gas density within the IGM, no such terms are necessary for modelling the forest power.

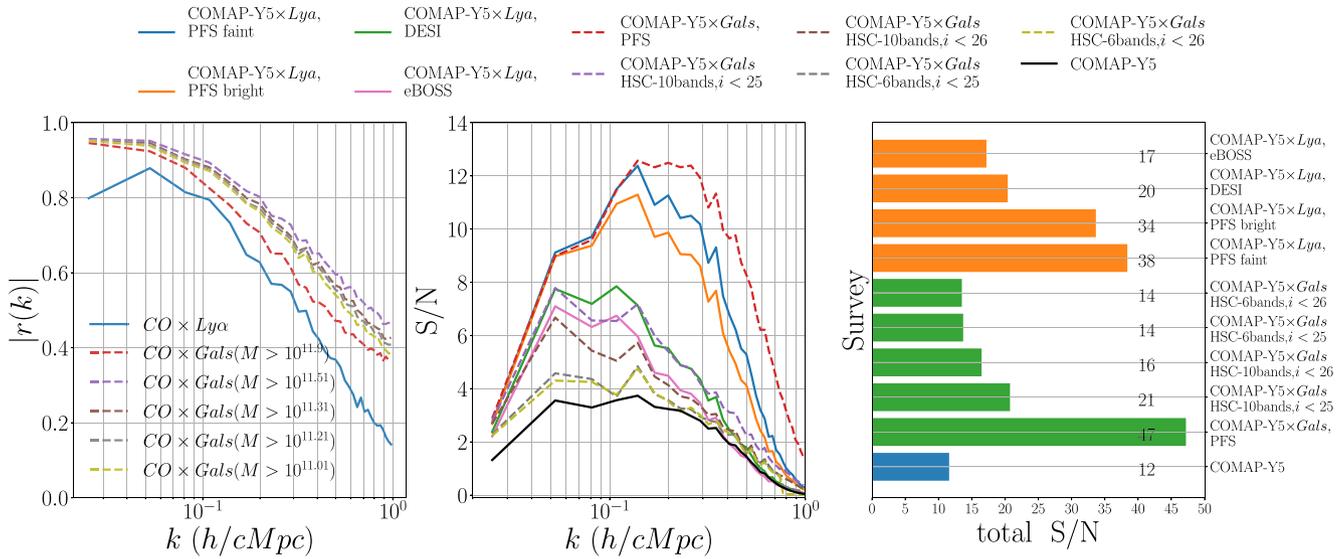

**Figure 1.** *Left-hand panel*: The plot shows the absolute values of the cross-correlation coefficients between auto CO power and CO × Ly-$\alpha$ or CO × galaxies power spectra. All signal maps are on a high-resolution grid of $(250\,h^{-1}\,\mathrm{ckpc})^3$. *Middle panel*: The S/N forecast of the realistic mock observations within the COMAP-Y5 Pathfinder's volume. The eBOSS observations are complete, while DESI data will be obtained in a 5-yr programme. The high-resolution PFS IGM map is planned to start observations in 2023. The spectroscopic galaxy survey is also planned with the PFS-GE programme and the photometric galaxy surveys, HSC + CLAUDES and HSC + CLAUDES + NIR, are either completed or will be completed as part of PFS-GE. Refer to Table 1 for a summary of the survey parameters. *Right-hand panel* illustrates the forecast total S/N for power over all $k$-bins, derived from equation (13). These values are shown on each bar for convenience.

A Gaussian likelihood is assumed for joint analysis in each scenario:

$$L \sim \frac{1}{N_k} \sum_i \sum_k G\left(P_i(k) - \hat{P}_i(k), \sigma_{P_i}(k)\right), \quad (20)$$

where $P_i(k)$ and $\sigma_{P_i}(k)$ are the simulated noiseless power spectrum and the estimated observational uncertainties, respectively (refer to Section 4). Here, $i$ iterates over all available autocorrelated and cross-correlated signals in each scenario. We draw Markov Chain Monte Carlo (MCMC) samples[4] from the posterior distribution obtained from the likelihood above, assuming flat priors.

The posterior constraints on all inferred parameters are shown in Fig. 2. In Fig. 3, the maximum a posteriori predictions for all modelled autopower and cross-power spectra are compared to the simulated signal. Due to the different mass completeness thresholds for the spectroscopic and photometric surveys, three sets of simulated galaxy power spectra are shown in the left-hand panel of Fig. 3. Tighter constraints on the model parameters from the Ly-$\alpha$ forest are observed, resulting in smaller uncertainties in the posterior power prediction compared to the galaxy surveys.

For galaxy surveys, we exclude non-linear scales (i.e. $k > 0.5\,h^{-1}\,\mathrm{cMpc}$) from the inference as they are not modelled well by the linear theory in equation (16). Removing the smallest scales is necessary to avoid a systematic offset in the parameter inference; nevertheless, the width of the posterior remains unchanged after this cut since the S/N is already small on small scales (refer to Fig. 1). The $k$-range of $[(1/250)\text{--}0.5]\,h\,\mathrm{Mpc}^{-1}$ adopted in our CO × galaxy analysis roughly matches the range constrained by the COMAP Early Science results (Chung et al. 2022). For the inference from the CO × Ly-$\alpha$ signal, however, we use the full $k$-range of $[(1/250)\text{--}1.0]\,h\,\mathrm{Mpc}^{-1}$.

---

[4]Using PYSTAN PYTHON package (Carpenter et al. 2017; Riddell, Hartikainen & Carter 2021).

Moreover, we find that there is an excess of Ly-$\alpha$ absorption signal on the largest scales, $k < 0.1\,h\,\mathrm{Mpc}^{-1}$. We attribute this to the He II reionization modelled in ASTRID, which enhances absorption on scales $L > 30\,h^{-1}\,\mathrm{cMpc}$. This leads to a larger correlation in the signal at $k < 0.1\,h\,\mathrm{Mpc}^{-1}$ (McQuinn et al. 2009, 2011; Gontcho A Gontcho et al. 2014; Pontzen 2014; Pontzen et al. 2014). The linear matter power model in equation (16) does not capture these effects in the Ly-$\alpha$ signal on large scales. Incorporating the model uncertainties into $\sigma_{P_{\mathrm{Ly}\alpha}}$ is not straightforward, so we do not attempt to do so in this work. Consequently, this model insufficiency results in discrepant inferred values for the Ly-$\alpha$ bias parameter, $b_{\mathrm{Ly}\alpha}$, across surveys with varying background source densities. We recommend using accurate simulation-based modelling, such as emulators, over biased linear perturbation for inferring from future observations. Emulator-based models (e.g. Fernandez, Ho & Bird 2022) offer a more reliable and precise approach to handling the complexities of large-scale structures. We further discuss the details of the Ly-$\alpha$ power spectrum in Appendix A.

## 6 DISCUSSION

Fig. 1 presents the predicted S/N for CO × Ly-$\alpha$ and CO × galaxies. The surveys with high spatial resolution, such as the Ly-$\alpha$ tomography surveys with dense background sources and medium-resolution spectroscopic galaxy redshift surveys planned with the upcoming Prime Focus multiplex Spectroscopy survey (PFS), are expected to enhance the detection sensitivity of COMAP-Y5 by approximately 200–300 per cent. Spectroscopic galaxy surveys are slightly more effective in enhancing the detection S/N on smaller scales, probably because galaxies are inherently better tracers of the line emission from the ISM. This is further supported by the cross-correlation coefficient presented in the left-hand panel of Fig. 1. We defer a thorough examination of this behaviour to future work.

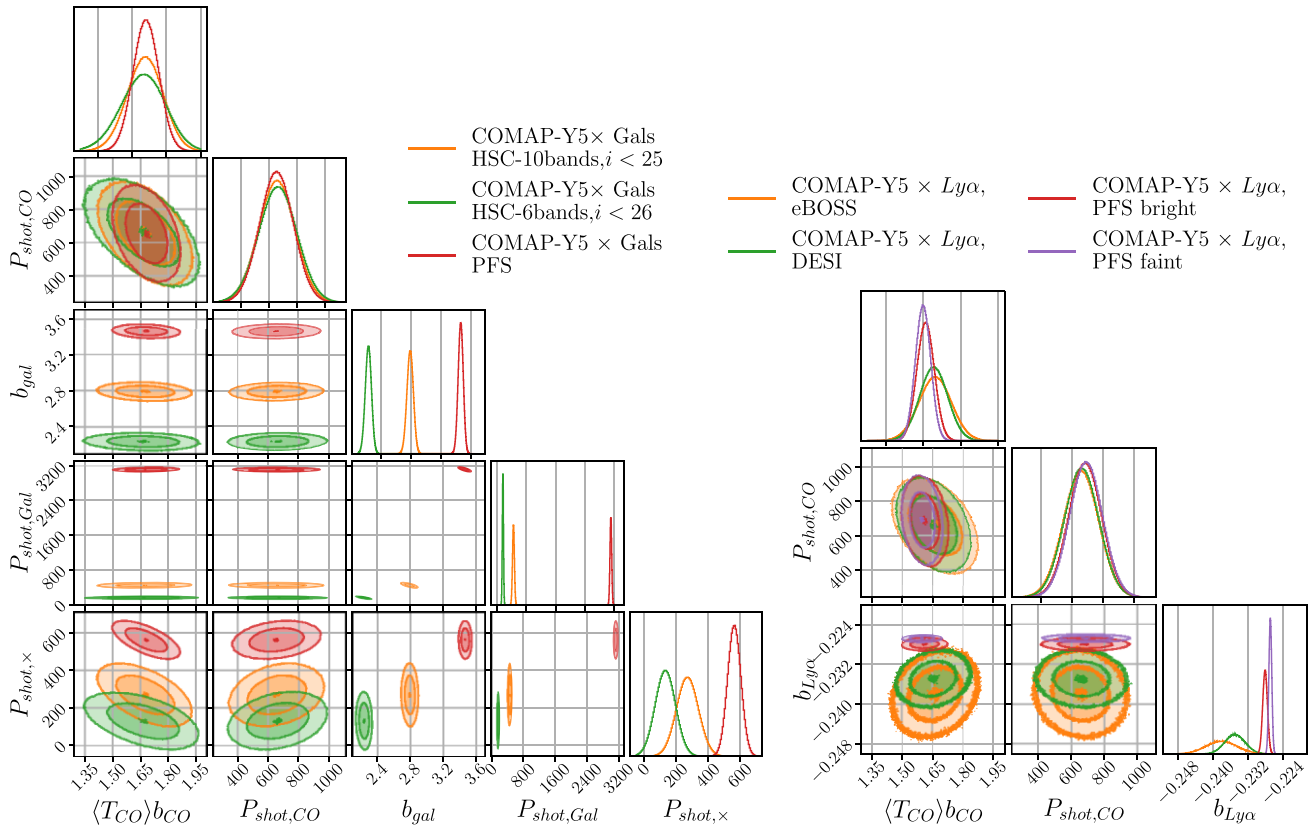

**Figure 2.** (*Left*) Forecasts for the inferred parameters in COMAP-Y5 × galaxy and (*Right*) COMAP-Y5 × Ly-α tomography (see Sections 5 and 6).

The tomography surveys with a lower background source density, such as the present-day eBOSS observations or the forthcoming DESI survey, improve the sensitivity by 50–75 per cent. On the other hand, a galaxy photometric catalogue, such as CLAUDS observed in $u + grizy$ bands (Sawicki et al. 2019), only marginally improves the detection sensitivity by ∼15 per cent. To match the S/N of COMAP-Y5 × galaxies to the CO × Ly-α forest in eBOSS or DESI, additional near-IR photometry in *YJHK* bands is required. Currently, $u + grizy + YJHK$ photometric observations are limited to smaller areas (see Table 1). Moreover, the CO emission signal is susceptible to large-scale contamination caused by foreground continuum emission from Milky Way, which weakens the cross-correlated S/N. This contamination increases the CO noise power, $\sigma_{P_{\rm CO,noise}}(k)$, on large parallel modes. The exact relevant scale where this contamination becomes prominent is not understood yet. The CO × photometric galaxy survey is, however, affected the most by foreground contamination as the signal originates mostly from the largest line-of-sight modes ($k_{\parallel,\rm min} < 0.02\,h\,\rm Mpc^{-1}$) due to the large galaxy redshift uncertainties. Fig. 4 shows the impact of this contamination on the forecast S/N, where the contamination is modelled by expanding the CO noise power, $P_{\rm CO,noise}$, on scales larger than $k_{\parallel,\rm min} = 0.01$ or $0.03\,h\,\rm Mpc^{-1}$ by a large factor of $10^9$. These scale cuts are conservative compared to those proposed by Ihle et al. (2022), which constrains the CO autopower on scales $k = 0.051$–$0.062\,h^{-1}\,\rm cMpc$.

We quantify the trade-off between finite sightline density and cosmic variance for the Ly-α forest observations by comparing the noise power spectrum and the signal for Ly-α tomography in Fig. 5, i.e. $P_{\rm Lya}(k)$ versus $P_{\rm Lya,n}(k)$. Our results show that for eBOSS and DESI, the primary factor affecting the cross-power spectrum

S/N at $k > 0.15\,h^{-1}\,\rm cMpc$ is the finite background source density. In contrast, for PFS forest observations, cosmic variance has the greatest impact at $k < 0.4$ or $0.5\,h^{-1}\,\rm cMpc$.

As detailed in Section 3.3, we adopt our fiducial CO emission model as the mean of the Gaussianized covariance provided by COMAP Early Science (Chung et al. 2022). However, this line emission model is still poorly constrained. To assess the sensitivity of our findings, we varied the model parameters within the $1\sigma$ range reported in table 5 of Chung et al. (2022). Interestingly, we observed that the rank ordering of the cross-power S/N forecast remained consistent with that of our fiducial emission model. Nevertheless, we noted that when the line emission is particularly strong, as seen with larger *C*, *A* or smaller *B*, *M*, the auto CO signal itself has a higher S/N, diminishing the utility of cross-correlation with coarse Ly-α tomographies or photometric galaxy surveys.

The EXCLAIM experiment (Switzer et al. 2021) will observe the [C II] emission at cosmic noon and is designed to overlap with BOSS Stripe 82, in order to benefit from [C II] × BOSS QSO sample (Pullen et al. 2022). However, on the scales where noise dominates over cosmic variance (refer to Fig. 5), we can compare the detection S/N of [C II] × QSO and [C II] × Ly-α forest data, that is eBOSS observations (Ravoux et al. 2020), as

$$\frac{(\rm S/N)_{C\,II\times Ly\alpha}}{(\rm S/N)_{C\,II\times QSO}} \sim \frac{b_{\rm Ly\alpha}}{b_{\rm QSO}} \left(\frac{P_{\rm QSO,noise}}{P_{\rm Ly\alpha,noise}}\right)^{1/2}. \quad (21)$$

Assuming a quasar bias factor and number density of $b_q = 3.64$ and $n_{\rm QSO} = 10^{-6}\,(h^{-1}\,\rm cMpc)^3$ (Font-Ribera et al. 2013; Eftekharzadeh et al. 2015), Ly-α bias factor of $b_{\rm Ly\alpha} = -0.20$ (Slosar et al. 2011), we find that the S/N of [C II] × Ly-$\alpha$ forest wins over cross-correlation against quasars by a factor of 10 or larger.

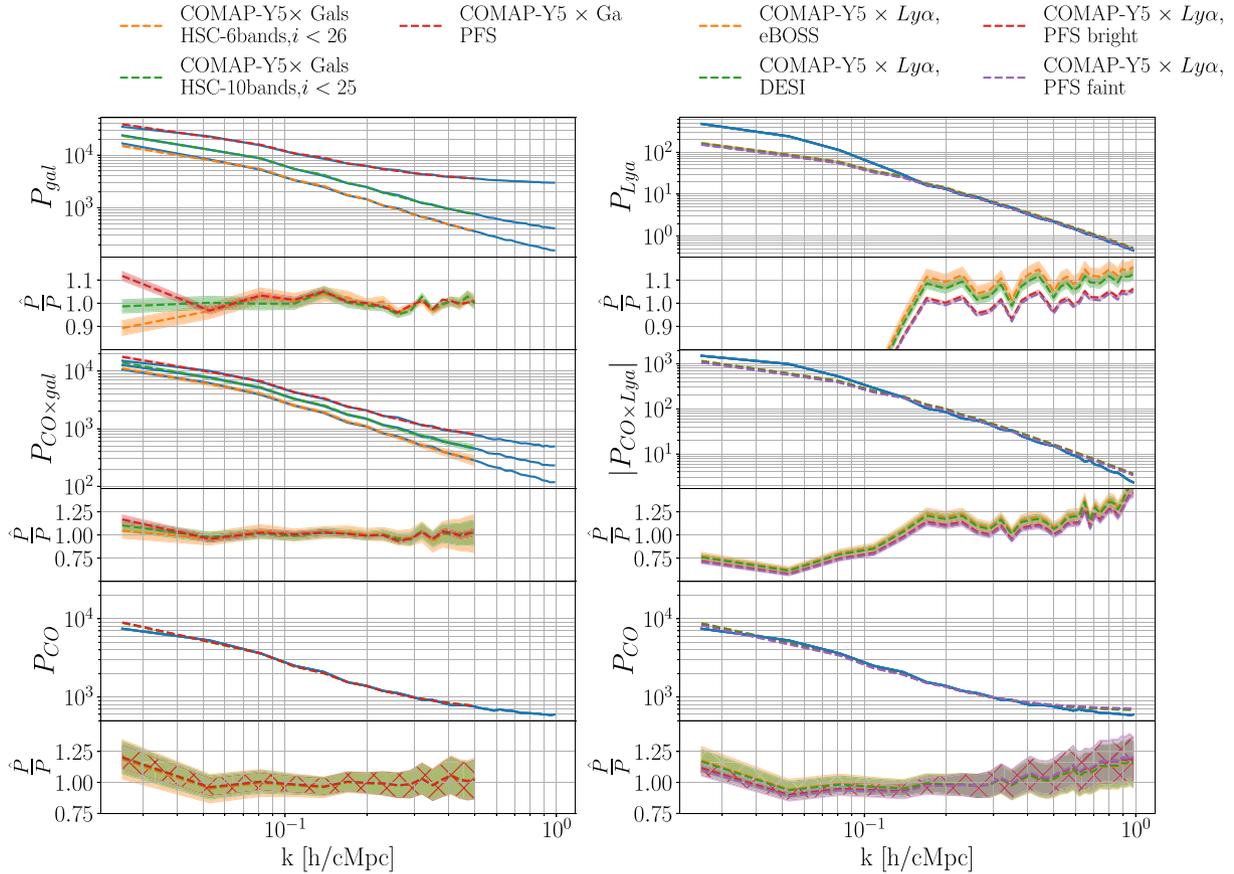

**Figure 3.** Comparison of the posterior prediction (dashed curves) and the signal power spectra (solid curves) with $1\sigma$ uncertainties indicated by shaded regions. The left-hand panel displays results for CO × galaxies with three separate curves for the power spectra signals due to varying mass completeness assumptions, which correspond to different magnitude cuts. The smallest scales, i.e. $k > 0.5\,h^{-1}$ cMpc, are excluded from the inference due to the linear model inadequacies in describing the signal on the smallest scales. The right-hand panel exhibits a significant deviation between the posterior prediction and signal for $P_{Ly\alpha}$ at $k < 0.1\,h^{-1}$ cMpc attributed to bubbles of enhanced H II fraction formed during the He II reionization. Further discussion is provided in Appendix A.

Croft et al. (2018) measure the Ly-$\alpha$ emission × Ly-$\alpha$ forest cross-correlation and the Ly-$\alpha$ emission × quasar cross-correlation function. The latter quantity is well detected, while the authors place an upper bound on the former cross-correlation. The upper bound limits the total Ly-$\alpha$ luminosity density from surrounding star-forming galaxies. On the other hand, Croft et al. (2018) suggest that the *detection* of diffuse Ly-$\alpha$ emission around quasars (mostly from relatively close to the quasars at 1–15 $h^{-1}$ cMpc) may largely result from reprocessed emission from the quasars themselves, rather than sourced by surrounding star-forming galaxies. On the other hand, Lin, Zheng & Cai (2022), a more recent reanalysis using SDSS DR16, proposes that the bulk of this diffuse Ly-$\alpha$ emission originates from the star-forming centre of galaxies or the diffuse gaseous haloes around them and not from the QSOs. Similarly, our work argues that the CO × Ly-$\alpha$ forest could provide insights into the origin of the CO emission, which has not yet been tightly constrained.

Clustering of the background sources in the Ly-$\alpha$ forest observations, such as quasars for eBOSS/DESI or LBGs for PFS surveys, increases the noise power spectrum, $P_{Ly\alpha,noise}$, by a factor of $1 + C_q(\boldsymbol{k}_\perp)n_{2D}$, where $C_q(\boldsymbol{k}_\perp)$ and $n_{2D}$ are the angular power spectrum and projected density of the sightlines contributing to the signal at a given redshift (McQuinn & White 2011). To investigate this issue, we measured the sightline clustering in the largest observed dense Ly-$\alpha$ tomography map, LATIS (Newman et al. 2020), which has a mean sightline separation comparable to PFS-faint. Our analysis reveals that the noise power spectrum in LATIS is only slightly elevated by a few per cent at $k = 0.03$–$0.1\,h$ Mpc$^{-1}$, the scales relevant to our study. This increase is expected to be even smaller for Ly-$\alpha$ surveys with lower sightline densities, such as eBOSS and DESI. As a result, the source clustering is not expected to affect our forecast for the S/N of the cross-correlated signal with CO emission.

The results from Section 5.2 are summarized in Fig. 6 and Table 2. The left-hand panel demonstrates that a PFS-like Ly-$\alpha$ tomography survey, with a transverse separation of $d_\perp = 2.5$–$3.7\,h^{-1}$ cMpc, is more effective at constraining the line emission power spectrum model than a spectroscopic galaxy survey planned with the same instrument, with $\sigma_z/(1+z) = 7 \times 10^{-4}$. The right-hand panel shows that coarser Ly-$\alpha$ tomography surveys, such as DESI with a larger mean sightline separation of $d_\perp = 10\,h^{-1}$ cMpc, provide better constraints than HSC-like photometric galaxy surveys observed in 10 bands ($u + grizy + YJHK$) with a redshift precision of $\sigma_z/(1+z) = 0.02$. These constraints are even comparable to those obtained from spectroscopic surveys. This is because the Ly-$\alpha$ power spectrum has a larger S/N and lower dimensionality of parameter space compared to galaxy surveys, where there are additional $P_{shot,gal}$ and $P_{\mathrm{shot},\times}$ parameters absent in the CO × Ly-α model, as shown in equation (16). Fig. 6 illustrates that *photometric* galaxy surveys offer only marginal improvements to the constraints from COMAP alone, which was expected based on the low forecast S/N shown in Fig. 1.

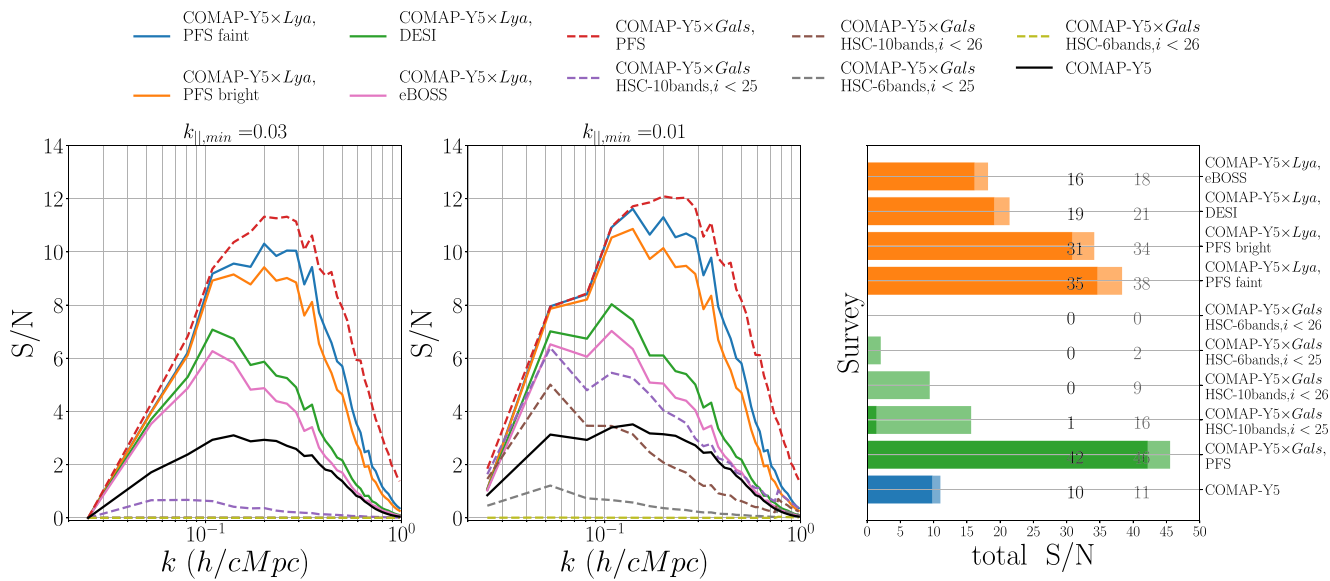

**Figure 4.** The forecast S/N, taking into account contamination from continuum foreground emission along the line of sight. Two different values of the minimum parallel wavenumber, $k_{\parallel,\mathrm{min}}$, are considered (0.01, 0.03) $h\,\mathrm{Mpc}^{-1}$, above which the CO power is assumed to be contaminated by interlopers. The exact $k$-cut is observationally unconstrained. In broad-band photometric galaxy surveys, only the parallel modes $k_{\parallel,\mathrm{min}} < 0.02\,h\,\mathrm{Mpc}^{-1}$ are typically measured, and these modes are the most contaminated by continuum foreground emission. The right-hand panel shows the total S/N, with different colour intensities indicating different values of $k_{\parallel,\mathrm{min}}$.

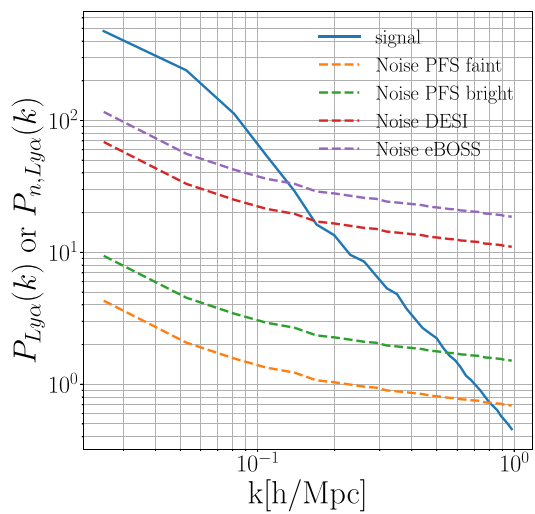

**Figure 5.** Comparing the impact of the cosmic variance and the finite background source density on the S/N of the Ly-α forest observations. For eBOSS and DESI, the S/N is primarily affected by the finite sightline density on scales $k > 0.15\,h^{-1}$ cMpc. Meanwhile, for PFS observations, the S/N is mostly dominated by cosmic variance at scales where $k < 0.4$ or $0.5\,h^{-1}$ cMpc.

The density of background sources in Ly-α forest maps and the accuracy of redshift estimations for galaxy surveys both decrease with increasing redshift. The eBOSS Ly-α forest survey has a mean transverse sightline separation of 8–16 $h^{-1}$ cMpc within the redshift range of $z = 2.2$–3.0 (Ravoux et al. 2020). The LATIS, which is similar to future PFS tomography surveys, has a mean sightline separation of 2.3–3.0 $h^{-1}$ cMpc within the redshift range of $z = 2.2$–2.8 (Newman et al. 2020). This study covers a range of sightline separations that falls broadly within this range. To compare with actual observed data, it is necessary to simulate the exact survey characteristics in the mock data, such as the mean source density in Ly-α forest observations or the redshift estimation errors for galaxy surveys.

In this work, we acknowledge that the auxiliary surveys (refer to Table 1) considered are assumed to fully overlap with the COMAP-Y5 volume, although this may not be the case for all observations. None the less, we believe that incorporating this factor is straightforward, and we hope that our forecast will inspire future decisions regarding observations.

## 7 SUMMARY

The LIM technique is a recent development for measuring the collective emission of specific atomic or molecular lines from galaxies of varying masses. However, detecting these faint signals still poses a challenge due to the necessary sensitivity. To improve the detection S/N, cross-correlation with other large-scale structure tracers has been found to be effective, such as combining LIM with galaxy redshift surveys or other LIM experiments. In this study, we explore a promising new survey method with exceptional spatial resolution, Ly-α tomography. Ly-α tomography uses a dense sample of background galaxies (or quasars) to create a 3D map of neutral hydrogen in the IGM (McQuinn & White 2011; Lee et al. 2014).

In particular, using large cosmological hydrodynamic simulations, we model the anticipated signal for the COMAP LIM experiment at the 5-yr benchmark (Section 3.3). Our findings are expected to apply to other molecular LIM experiments with similar instrumental noise; however, one should adopt the appropriate model for the emission line observed in each particular LIM survey. We also made mock observations of the Ly-$\alpha$ absorption signal for fully observed tomography surveys, such as eBOSS (Ravoux et al. 2020), and those expected to be completed in the coming years, such as DESI (Chaussidon et al. 2023 ) or PFS (Greene et al. 2022) (Section 3.1). The key variable for Ly-$\alpha$ tomography surveys is the mean separation between the observed background sources. The map area covered by these surveys is comparable to or exceeds the coverage for COMAP-Y5. In this work, we assume that these auxiliary maps fully overlap with the observed volume of COMAP-Y5. For a comprehensive list of survey parameters, including the volume coverage, please consult Table 1.

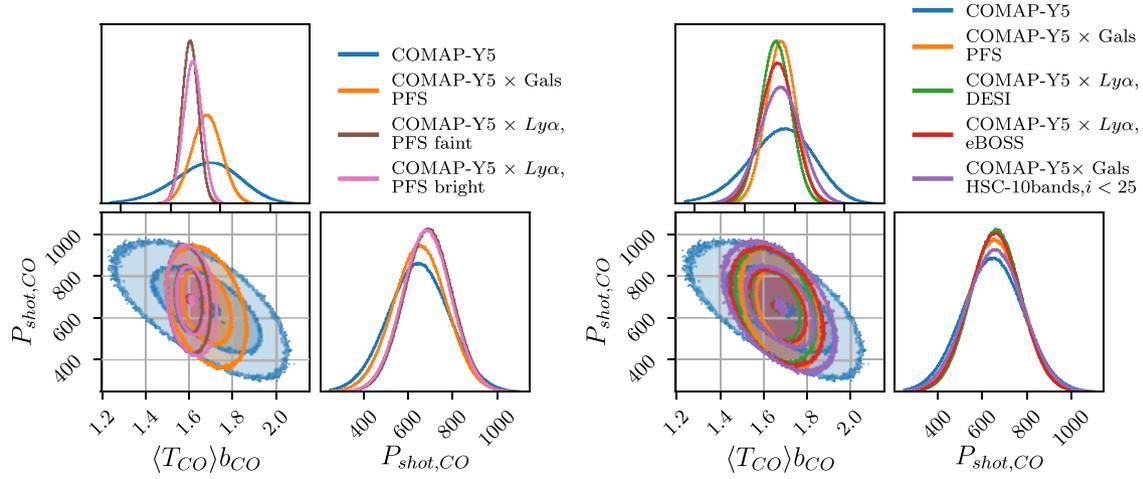

**Figure 6.** The constraints on CO bias and shot noise in the linear bias power spectrum formalism. Joint analyses, including both autopower and cross-power spectra, provide tighter constraints. The left-hand panel demonstrates that the constraints from CO × a PFS-like Ly-$\alpha$ tomography survey are tighter compared to a CO × PFS-like spectroscopy survey (refer to Section 6 for more details). The right-hand panel shows that CO × an eBOSS- or DESI-like Ly-$\alpha$ survey yields tighter constraints than CO × HSC-like photometric galaxy surveys. The medians and standard deviations of these distributions are tabulated in Table 2. Also, refer to Fig. 2 for the posteriors on the complete set of parameters.

**Table 2.** Comparison of forecast inference results for CO emission clustering across different surveys. The complete posterior distributions are shown in Figs 2 and 6. Physical units are in μK and μK$^2$(cMpc $h^{-1}$)$^3$ for $\langle T_{\rm CO}\rangle b_{\rm CO}$ and $P_{\rm shot,CO}$, respectively.

| Survey | $\langle T_{\rm CO}\rangle b_{\rm CO}$ | $P_{\rm shot,CO}$ |
| --- | --- | --- |
| COMAP-Y5 | 1.68 ± 0.17 | 647 ± 136 |
| COMAP-Y5 × Gals HSC – 6 bands, $i < 26$ | 1.67 ± 0.13 | 663 ± 135 |
| COMAP-Y5 × Gals HSC – 10 bands, $i < 25$ | 1.68 ± 0.10 | 657 ± 128 |
| COMAP-Y5 × Gals PFS | 1.68 ± 0.08 | 654 ± 119 |
| COMAP-Y5 × Ly$\alpha$, eBOSS | 1.66 ± 0.09 | 657 ± 114 |
| COMAP-Y5 × Ly$\alpha$, DESI | 1.65 ± 0.08 | 662 ± 112 |
| COMAP-Y5 × Ly$\alpha$, PFS-bright | 1.61 ± 0.05 | 684 ± 108 |
| COMAP-Y5 × Ly$\alpha$, PFS-faint | 1.60 ± 0.04 | 691 ± 107 |

The findings of this study highlight the potential benefits of utilizing the Ly-$\alpha$ forest to aid in the initial detection of signals in line intensity experiments. The enhancement of the S/N for the cross-correlated CO emission with any auxiliary survey depends on the spatial resolution and the noise in the auxiliary data (Chung et al. 2019). The cross-correlation between COMAP-Y5 and the PFS Ly-$\alpha$ tomography survey will enhance the detection S/N by ∼200–300 per cent, comparable to medium-resolution spectroscopic galaxy surveys planned with the same instrument (Fig. 1). The cross-correlation signal with sparser Ly-$\alpha$ tomography surveys, such as eBOSS and DESI, still enhances the detection S/N by 50–75 per cent.

Our results can be readily applied to actual existing data thanks to the observed quasar spectra in eBOSS Stripe 82, which covers an extensive area of 220 deg$^2$. Additionally, we demonstrate that the clustering of CO emission sources can be tightly constrained by the Ly-$\alpha$ tomography surveys. This is possible as a result of the elevated S/N in the cross-correlation, as well as the uncomplicated nature of Ly-$\alpha$ absorption power spectrum modelling compared to the galaxy redshift surveys in CO × galaxies. However, a joint constraint on the emission clustering from CO × Ly-$\alpha$ and CO × galaxies is expected to provide further consistency tests on the inferred parameters. Our findings are presented in Sections 5 and 6, depicted in Fig. 6, and summarized in Table 2.

It should be noted that any foreground contamination, or other systematics, that are unique to CO and Ly-$\alpha$ forest surveys, will not lead to biases in the inferences from the cross-spectrum signal. For example, residual foreground contamination might strongly bias a CO autopower spectrum but will not lead to a spurious correlation with the Ly-$\alpha$ forest on average.

Section 6 presents a simple order-of-magnitude calculation, which suggests that the cross-correlation with Ly-$\alpha$ forest fluctuations would also be beneficial for the EXCLAIM survey (Switzer et al. 2021). EXCLAIM will observe [C II] (1900-GHz rest-frame) emissions at cosmic noon and has significant overlap with BOSS quasars (Pullen et al. 2022).

In Section 6 and Appendix A, we emphasize that precise modelling of the galaxy line emission and the Ly-$\alpha$ absorption signals is necessary for an accurate inference from actual data. Consequently, we defer a thorough emulator-based inference using cosmological simulations similar to those in previous studies (Bird et al. 2019; Fernandez et al. 2022; Ho, Bird & Shelton 2022) to future work.

## ACKNOWLEDGEMENTS

MQ was supported by National Science Foundation grant AST-2107821. SB acknowledges funding support from National Aeronautics and Space Administration (NASA) grant NASA-80NSSC21K1840. The authors acknowledge the Frontera computing project at the Texas Advanced Computing Center (TACC) for providing High-Performance Computing and storage resources that have contributed to the research results reported within this paper. Frontera is made possible by National Science Foundation award OAC-1818253 (http://www.tacc.utexas.edu). We especially thank Ming-Feng Ho, Martin Fernandez, and Reza Mondai for all the constructive comments.

## DATA AVAILABILITY

The ASTRID simulation snapshots utilized in this research are accessible upon request. Additionally, the analysis scripts, cookbook notebooks, and data generated during this study are all accessible on our GitHub repository at https://github.com/qezlou/lila.

# APPENDIX A: ROBUSTNESS OF THE RESULTS TO SIMULATION CHOICE

To verify the robustness of our results to the cosmological simulation used, we conduct the forecast analysis using a second cosmological hydrodynamic simulation, TNG300-1 (Nelson et al. 2019). ASTRID has 1.8 times larger volume and increased particle mass resolution, as well as models for patchy hydrogen and helium reionization. Inference on the mock signal generated from the TNG300 simulation is presented in Fig. A1. The posteriors on $\langle T_{CO}\rangle b_{CO}$ and $P_{\rm shot,CO}$ and the rank order of the S/N predictions for CO × Ly-α and CO × galaxies have remained unchanged. Unlike the results shown in Fig. 2, the inferred $b_{Ly\alpha}$ from multiple surveys mocked using TNG300 are now in agreement. This agreement can be attributed to the following.

The difference between TNG300 and ASTRID is because TNG300 does not incorporate a model for He II reionization, which boosts power on scales comparable to the size of the reionized bubble (30 $h^{-1}$ cMpc). In ASTRID, guided by the radiative transfer simulations of McQuinn & White (2011), He II reionization is modelled by the creation of 30 $h^{-1}$ cMpc ionized bubbles around potentially quasar-hosting haloes (Upton Sanderbeck & Bird 2020). Comparing the 3D autopower spectrum of the Ly-α absorption signal in ASTRID with TNG300-1 in Fig. A2 shows that ASTRID predicts roughly an order of magnitude larger power at scales $k < 0.1\,h\,\text{Mpc}^{-1}$. This large-scale power enhancement is expected in any patchy reionization model, as discussed in Gontcho A Gontcho et al. (2014), Pontzen (2014), and Pontzen et al. (2014).

However, patchy reionization is not included in either the linear bias model outlined in equation (16), or TNG300, which thus agree with each other, although neither would agree with the true expected signal.

Damped Ly-α absorbers (DLAs) and Lyman limit systems (LLSs) produce substantial absorption in the spectrum. The effect of the damping wings in these absorbers resembles the observed overshooting of power on large scales (Rogers et al. 2018). The contribution of DLAs/LLSs to the power spectrum can be reduced by masking these absorbers, as is also done in observational surveys (Newman et al. 2020). Fig. A2 shows that there is still a power spectrum excess on large scales, even after masking all the absorbers with equivalent width (EW) >5 Å. This indicates that DLAs/LLSs alone cannot fully account for the excessive power at large scales.

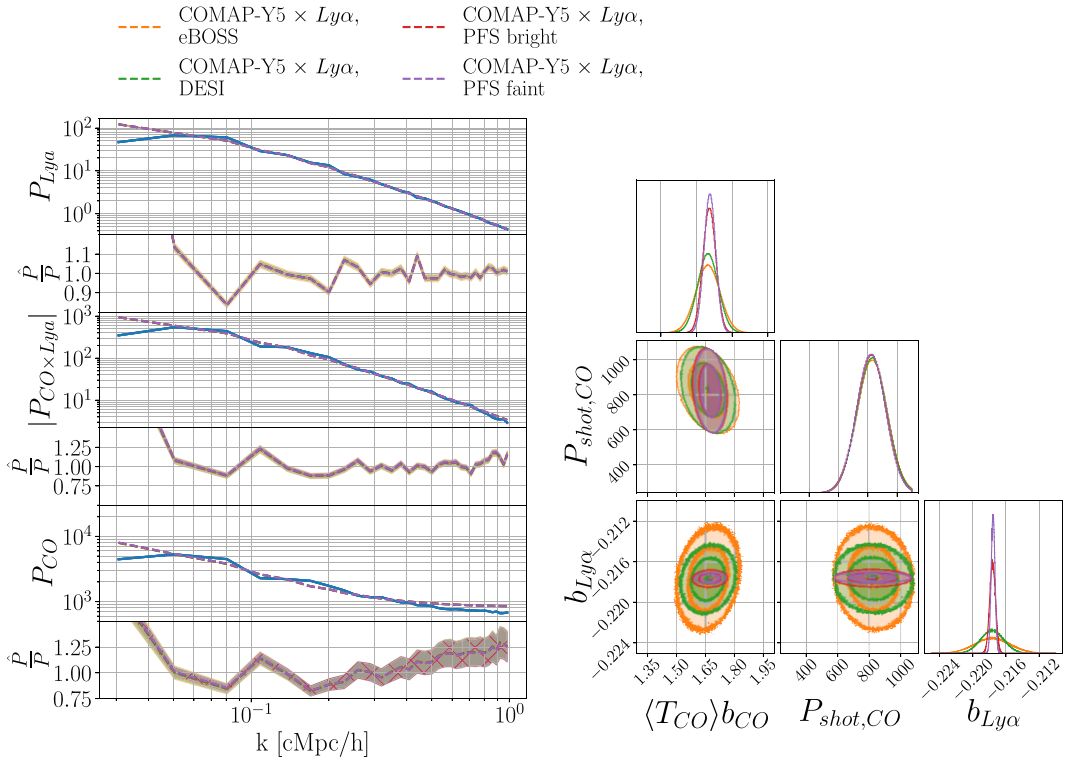

**Figure A1.** *Validation with an alternate cosmological simulation*: (*Left*) The maximum a posteriori power spectra. (*Right*) The inferred parameters for COMAP-Y × Ly-α tomography mock observations from the TNG300 hydrodynamical simulation. Compared to Fig. 2, consistency in the inferred $b_{Ly\alpha}$ values is seen among surveys. The absence of He II reionization in the TNG300 simulation allows a biased linear power spectrum to fit the largest scales well.

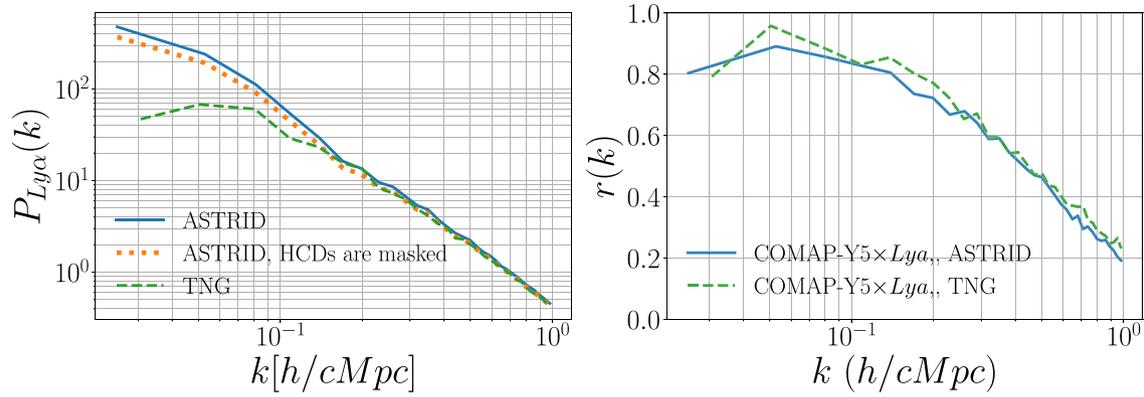

**Figure A2.** This figure demonstrates the effect of He II reionization on the largest scales of the Ly-α power spectrum. *Left*: A higher power at $k < 0.1\,h\,\mathrm{Mpc}^{-1}$ is observed in ASTRID compared to TNG300. Masking the DLAs/LLSs with EW > 5 Å in ASTRID (dotted curve) does not substantially reduce the power on large scales. Thus, the excess power is due to the presence of ionized helium bubbles formed during He II reionization. *Right*: The large-scale suppression of the cross-correlation coefficient due to He II reionization is evident for ASTRID.

This paper has been typeset from a TEX/LATEX file prepared by the author.